\documentclass[pre,twocolumn,groupedaddress,showpacs,showkeys,amsmath]{revtex4}

\usepackage{graphicx}
\usepackage{dcolumn}
\usepackage{bm}

\newcommand{\change}[1]{{\bf #1}}

\begin{document}


\title{Information theoretic approach to ground-state phase transitions for two and three-dimensional frustrated spin systems}

\author{O. Melchert}
\email{oliver.melchert@uni-oldenburg.de}
\author{A. K. Hartmann}
\email{alexander.hartmann@uni-oldenburg.de}
\affiliation{
Institut f\"ur Physik, Universit\"at Oldenburg, Carl-von-Ossietzky Strasse, 26111 Oldenburg, Germany\\
}

\date{\today}


\begin{abstract}
The information theoretic observables entropy  (a measure of
disorder), excess entropy (a measure of complexity) and  multi
information  are used to analyze ground-state spin configurations for
disordered and frustrated model systems in $2D$ and $3D$.  For both
model systems, ground-state spin configurations can be obtained in
polynomial time  via exact combinatorial optimization algorithms,
which allowed us to study large systems with high numerical accuracy. 
Both model systems exhibit a continuous transition from an ordered to
a  disordered ground state as a model parameter is varied.  By using
the above information theoretic observables it is possible to detect
changes in the spatial structure of the ground states as the critical
point is approached.  It is further possible to quantify the scaling
behavior of the information theoretic observables in the vicinity of
the critical point.  For both model systems considered, the estimates
of critical properties  for the ground-state phase transitions are in
good agreement with  existing results reported in the literature.
\end{abstract} 

\pacs{75.40.Mg, 05.70.Jk}
\maketitle

\section{Introduction}
\label{sect:introduction}

The standard analysis of physical phase transitions
involves the analysis of order parameters and other derivatives of
the free energy \cite{goldenfeld1992}. An alternative approach is based 
not on physical but information-theoretic approaches, an approach which
has recently started to be occasionally used in the analysis of (more or less) 
complex systems \cite{crutchfield2010}.

The presented study extends on previous studies that 
employed information-theoretic methods to measure entropy
(i.e.\ disorder and randomness) and statistical complexity 
(i.e.\ structure, patterns and correlations) for
$d\geq1$-dimensional systems 
\cite{Arnold1996,Crutchfield1997,Feldman2003,Feldman2008}.
For $1D$ systems, the excess entropy constitutes a well understood 
information-theoretic measure of complexity. 
Effectively, it accounts for the rapidity of entropy convergence.
While the extension of entropy to higher dimensions is rather 
intuitive, the extension of excess entropy is not. 
As a remedy, in Ref.\ \cite{Feldman2003}, three different approaches 
were developed in order to extend the definition of excess entropy to $d>1$,
allowing to quantify the complexity for spatial systems 
in higher dimensions.

Most previous studies  focused on 
characterizing the above information-theoretic observables for 
systems without disorder, only. 
As regards this, the local states method \cite{Schlijper1990}, proposed 
for the calculation of free energies within importance-sampling 
Monte Carlo (MC) simulations, was based on entropy estimation techniques 
for lattice models with discrete interactions and translation-invariant 
interactions (i.e.\ non-disordered, pure systems). 
It combined upper and lower bounds for the entropy density 
to compute free energies (along with an error estimate). 
A comparison of numerical simulations to exact results for the $2D$ Ising 
ferromagnet indicated that it yields reliable estimates already for short 
simulation runs (even in the critical region). 

Using quite similar entropy estimation techniques, the simulations reported 
in Ref.\ \cite{Feldman2003} were performed for the 
$2D$ square lattice Ising model with nearest and next nearest neighbor interactions.
By means of single spin-flip Metropolis dynamics at a comparatively low temperature,
two variants of the excess entropy were put under scrutiny. 
A careful analysis indicated that these are sensitive to changes in the 
spatial structure of the spin configurations as the nearest neighbor 
coupling strength was varied.
They were further found to be superior to conventional structure factors.
This study allows to conclude that the excess entropy in 
$2D$ comprises a general purpose measures of $2D$ structure.

Only recently, Ref.\ \cite{Robinson2011} used similar methods to characterize 
local, i.e.\ lattice site dependent, entropies and local excess entropies for 
the Kaya-Berker model. The latter is based on the Ising antiferromagnet (IAFM) 
on a triangular lattice, wherein a particular sublattice is diluted, only.
The IAFM exhibits geometric frustration and does not order at 
finite temperature. In contrast to the pure model, the Kaya-Berker model orders 
at finite temperature if at least a fraction $p=0.0975$ of the sites on the 
diluted sublattice are deleted.
The simulations were performed using single spin-flip Metropolis dynamics 
at fixed dilution $p=0.15$, where the freezing temperature amounts to 
$T_c\approx 0.84$. In the simulation, various temperatures down to $T=0.4$
were considered.
It was found that the distribution of local entropies broadens in the glassy 
phase below $T\approx 0.8$, indicating that for low temperatures local entropy 
is not homogeneously distributed over the lattice.
Further, the average of the local excess entropy was observed to exhibit a 
pronounced peak at the critical temperature, indicating that it is sensitive 
to structural changes for the $2D$ configurations as a result of the spin 
glass ordering. Finally, complexity-entropy diagrams for the frustrated 
Kaya-Berker model, recorded at various temperatures, were found to be 
qualitatively different from those corresponding to pure models 
(see Ref.\ \cite{Feldman2008}). 
This study allows to conclude that local entropy density and local excess 
entropy are valuable observables that yield insight to local structure and 
randomness for frustrated $2D$ systems.

\begin{figure}[t!]
\begin{center}
\includegraphics[width=1.0\linewidth]{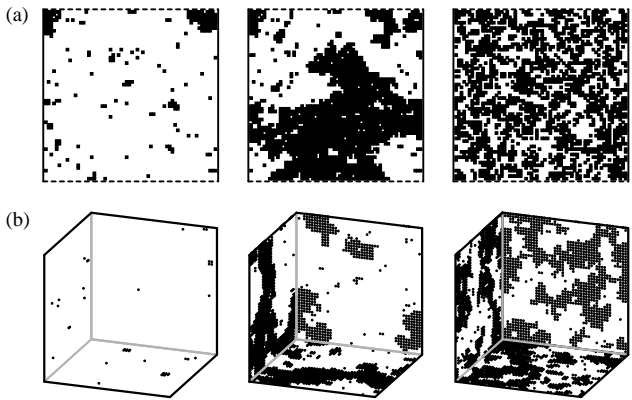}
\end{center}
\caption{\label{fig:samples}
GS samples for 
(a) the $L=64$ $2D$ RBIM with mixed boundary conditions (periodic = solid line, free = dashed line) at $\mu=1.15,~0.97,~0.40$ (from left to right), and
(b) the $L=48$ $3D$ RFIM with fully periodic boundary conditions at $b=2.0,~2.4,~2.9$ (from left to right; only the spins on the $x,y,z=0$ planes are displayed).
}
\end{figure}

Here, we aim to characterize the spatial structure displayed by 
exact ground states (GSs) of disordered model systems in terms of the 
information-theoretic observables entropy and excess entropy. 
For this purpose we consider the $2D$ random bond Ising model ($2D$ RBIM) 
as well as the $3D$ random field Ising model ($3D$ RFIM),
which both exhibit disorder-driven zero-temperature
phase transitions.
More precise, we use one of the approaches developed in Ref.\ \cite{Feldman2003},
aimed at understanding spatial patterns for $2D$ systems by parsing them 
into $1D$ sequences. 
In addition we also consider the multi-information, which, for 
the ordinary Ising ferromagnet in the thermodynamic limit was 
proved to be maximized at the critical point \cite{Erb2004}.
In contrast to the previous studies we work at $T=0$, aiming to characterize 
the ground-state phase transitions that appear as the disorder is varied 
for the above two disordered model systems (see sect.\ \ref{sect:models}).
In doing so, we were interested whether one can observe a change in the structure
of the ground states by using the above information-theory inspired observables.
Further, if the latter is possible, it is of interest to quantify the scaling behavior
of these observables in the vicinity of the phase transition. 

The remainder of the presented article is organized as follows.
In section \ref{sect:models} we introduce the spin models that are
considered in the presented study.
In section \ref{sect:infTheor}, the information-theoretic
observables entropy and excess entropy are introduced and illustrated 
in more detail. Section \ref{sect:results} reports on the numerical results,
further discussed in section \ref{sect:discussion}. 
An elaborate summary of the presented article is available at the {\emph{papercore database}}
\cite{papercore}.


\section{Models}
\label{sect:models}

Next, we present the two models which were studied in this work,
the $2D$ RBIM and the $3D$ RFIM.

\subsection{$2D$ random bond Ising model}
\label{subsect:2DRBIM}

We investigated GSs for the $2D$ RBIM, considering 
a square lattice of side length $L$.
The respective model consists of $N=L^2$ Ising spins, 
for which a particular spin configuration might be written as
$\sigma=(\sigma_1,\ldots,\sigma_N)$, where $\sigma_i\in \{+1,-1\}$. 
The energy of a given spin
configuration is measured by the Edwards-Anderson Hamiltonian 
\cite{edwards1975}
\begin{eqnarray}
H_{\rm RBIM}(\sigma) = -\sum_{\langle i,j \rangle} J_{ij} ~\sigma_i \sigma_j,
\label{eq:EA_hamiltonian}
\end{eqnarray}
where the sum is understood to run over all pairs of nearest-neighbor spins
(on a $2D$ square lattice),
with periodic boundary conditions (BCs) in the $x$-direction and free BCs 
	in the $y$-direction.  
In the above energy function, the bonds $J_{ij}$ are quenched random variables 
drawn from the disorder distribution
\begin{align}
P(J_{ij})= \exp[ -(J_{ij}-\mu)^2/(2 \sigma_J^2)]/(\sigma_J \sqrt{2 \pi}),
\end{align}
where the width of the distribution was fixed to $\sigma_J=1$.
Consequently, one realization of the disorder consists of a mixture of 
antiferromagnetic bonds ($J_{ij}<0$) that prefer an antiparallel alignment of 
the coupled spins, and ferromagnetic bonds ($J_{ij}>0$) in 
favor of parallel aligned spins.
In general, the competitive nature of these interactions gives rise to frustration. 
A plaquette, i.e.\ an elementary square on the lattice, is said to be frustrated if it is
bordered by an odd number of antiferromagnetic bonds. 
In effect, frustration rules out a GS in which all the bonds are satisfied. 
As limiting cases one can identify the random bond Ising ferromagnet (FM) 
\change{as $\mu\to \infty$} and 
the (Gaussian) $2D$ Edwards Anderson Ising spin glass (SG)
\cite{edwards1975,SG2dReview2007,Hartmann2DSG2011} at $\mu=0$. 
A GS spin configuration  $\sigma_{GS}$ 
is simply a minimizer of the energy function 
Eq.\ \ref{eq:EA_hamiltonian}. 
Thus, regarding the GSs as a function of the variable $\mu$, we expect to find 
a ferromagnetic phase (spin glass phase) for $\mu>\mu_c$ ($\mu<\mu_c$) 
	wherein $\mu_c$ denotes the critical point at which the $T\!=\!0$ FM-SG transition, i.e.\
a continuous disorder-driven phase transition, takes place.
Samples of GSs for the $2D$ RBIM for different values of the parameter $\mu$ are 
shown in Fig.\ \ref{fig:samples}(a).

The Ising spin glass is a paradigmatic model for a disordered magnet. Since the effects of 
the disorder are well visible at zero temperature, the investigation of ground 
state properties is of prime importance. 
For a planar version of this model, e.g., a $2D$ square lattice with periodic 
boundary conditions in only one direction,
a solution of the GS problem is possible by means of a mapping to an 
appropriate minimum-weight perfect-matching problem. 
This latter problem can be solved by means of exact combinatorial optimization 
algorithms from computer science
\cite{SG2dReview2007,opt-phys2001,bieche1980,barahona1982,thomas2007,pardella2008}, whose
running time increases only polynomially with the system size. Hence, very
large systems can be treated exactly, giving very precise and reliable
estimates for the observables.  
The GS problem for lattice dimensions $d\!>\!2$ or systems subject
to an external magnetic field belong to the class of nondeterministic 
polynomial (NP)-hard problems \cite{bieche1980,cipra2000}. For those problems, no exact algorithm with 
a polynomial running time has been found so far.
From a conceptual point of view, 
the existence of numerically exact and highly efficient
algorithms for the $2D$ SG with periodic 
boundary conditions in at most one direction motivates the special interest 
in this setup during the last decades. 

As pointed out above, for the $2D$ RBIM on planar lattice graphs (including the Ising spin glass),
GS spin configurations can be found in polynomial time. 
We here use a particular mapping to an appropriate minimum-weight perfect matching problem, 
presented in Ref.\ \cite{pardella2008}. The use of this approach permits the treatment of large
systems, easily up to $L\!=\!512$, on single processor systems.  
In a previous study \cite{Melchert2009}, we performed such GS calculations 
and employed a finite-size scaling analysis for systems of moderate sizes 
($L\!\leq\!64$) to locate the critical point at which the transitions takes place.  
Therefore, we analyzed the Binder parameter \cite{binder1981} 
$b_L\!=\!\frac{1}{2} [3- \langle m_L^4\rangle / \langle m_L^2\rangle^2]$
that is associated to the magnetization $m_L$ and is expected to scale as
$b_L(\rho)\!\sim\!f[ (\rho-\rho_{\rm c}) L^{1/\nu}]$,
where $f$ is a size-independent function and $\nu$ signifies the critical 
exponent that describes the divergence of the correlation length as the 
critical point is approached. 
The magnetization is simply the sum of all spin-values in the GS spin configuration, i.e.\
$m_L=\sum_i \sigma_{{\rm GS} i}/L^{2}$.
Using the data collapse anticipated by the scaling assumption above we obtained
$\mu_{\rm{c}}^{\rm lit}\!=\!1.031(2)$ and $\nu^{\rm lit}\!=\!1.49(4)$ (for a further study of this
transition, yielding similar results using renormalization group techniques, 
see Ref.\ \cite{McMillan1984}).
Finally, we characterized the transition using a finite-size scaling analysis 
for the largest and second-largest ferromagnetic clusters of spins within the 
GS spin configurations. These clusters form as one proceeds from 
spin glass ordered to ferromagnetic ground states.
The results obtained from the related scaling analysis support the results
obtained earlier by considering the Binder parameter and magnetization.

\subsection{$3D$ random field Ising model}
\label{subsect:3DRFIM}

We further investigated GSs for the 
$3D$ RFIM on a simple cubic lattice of side length $L$.
The respective model consists of $N=L^3$ Ising spins, 
and the energy of a given spin
configuration is measured by the Hamiltonian
\begin{eqnarray}
H_{\rm RFIM}(\sigma) = -J \sum_{\langle i,j \rangle} ~\sigma_i \sigma_j - \sum_i b_i\sigma_i,
\label{eq:RFIM_hamiltonian}
\end{eqnarray}
where the first sum is understood to run over all pairs of nearest-neighbor spins
on a $3D$ simple cubic lattice with fully periodic BCs.
In the above energy function, the bonds that couple adjacent spins are ferromagnetic, i.e.\ $J>0$,
and the local fields $b_i$ ($i=1,\ldots,N$) introduce disorder to the model.
The values of the local fields are independently drawn from the disorder distribution
\begin{align}
P(b_i)= \exp[ -(b_i)^2/(2 b^2)]/(b \sqrt{2 \pi}),
\end{align}
where the mean of the distribution is fixed to zero and the width amounts to $b$.
Thus, one realization of the disorder consists of ferromagnetic spin-spin couplings with each spin coupled
to a local random field.

The RFIM is a basic model for random systems \cite{fischer1991,young1998} and also gives rise to frustration. 
While in order to minimize Eq.\ \ref{eq:RFIM_hamiltonian} 
the ferromagnetic spin-spin coupling will tend to align coupled spins in parallel,
the random fields will tend to align the spins parallel to the local field, possibly introducing 
a paramagnetic effect on the GSs.
For a field width that is small compared to the ferromagnetic coupling, i.e.\ for $b\ll J$, one 
might expect a dominance of the ferromagnetic spin-spin coupling in the GS spin configurations. 
In contrast to this, for a comparatively large field width, the orientation of the majority of spins in a GS
will be determined by the local random fields, suggesting a paramagnetic GS for large values of $b$.
As a result, at $T=0$, the RFIM exhibits a disorder driven, continuous ferromagnet to paramagnet (PM) 
transition regarding the ground state structure at a finite value $b_{\rm c}$.
Samples of GSs for the $3D$ RFIM for different values of the field strength $b$ are 
shown in Fig.\ \ref{fig:samples}(b).

The solution of the GS problem for the general RFIM (not only its $3D$ variant) is 
possible by means of a mapping to an appropriate maximum-flow problem \cite{bastea1999,opt-phys2001,rieger2002}. 
This latter problem can be solved in polynomial time by means of exact combinatorial optimization 
algorithms from computer science.
As for the $2D$ RBIM this offers the possibility to study 
large systems, easily up to $L\!=\!64$, giving very precise and reliable estimates for the observables.  

Using such optimization methods, the critical point and critical exponents for the GS phase transition
can be obtained by analyzing the magnetic properties of the model \cite{hartmann1998,dAuriac1997,hartmann2001}. 
For the $3D$ RFIM we here
quote the values $b_{\rm c}^{\rm lit}=2.28(1)$ for the critical point and $\nu^{\rm lit}=1.32(7)$ for the critical exponent 
that describes the divergence of the correlation length as the critical point is approached.

Further, the $3D$ RFIM also exhibits a particular percolation transition. I.e., regarding the 
simultaneous spanning of up and down spin domains as a function of the field strength $b$,
Ref.\ \cite{Seppala2002} report on the percolation critical point $b_{\rm p}^{\rm lit}= 2.32(1)$ 
and the value $\nu_{\rm p}^{\rm lit}=1.00(5)$ (note that for random percolation in $3D$ one has 
$\nu_{\rm p}=0.88$; see Ref.\ \cite{stauffer1994}).
However, the results appeared to depend on the particular spanning criterion. In this regard, in 
the limit of the spanning probability approaching to zero (indicating the onset of percolation)
the scaling analysis extrapolated to $\nu_{\rm p}^{\rm lit}=1.3(1)$.


\section{Information-theoretic observables in $1D$ and extension to $d=2$ and $3$}
\label{sect:infTheor}

In the current section we introduce basic notations from information theory,
needed to define the entropy rate, excess entropy, and multi information that might
be associated to a ($1D$) sequence of symbols. 
In this regard, for the definition of the entropy rate and excess entropy we 
follow the notation of Refs.\ \cite{Shalizi2001,Crutchfield2003,Feldman2008}. For the 
definition of the multi information we follow Ref.\ \cite{Erb2004}.
In the above references, a more elaborate discussion of the individual 
information-theoretic observables can be found.

A prerequisite for the definition of these observables is
the $M$-block \emph{Shannon entropy} $H[S^M]$ for a block of $M$ consecutive 
random variables $S^M=S_1\ldots S_M$. Each random variable $S_i$ might assume 
values $\sigma_i\in\mathcal{A}$, where the set $\mathcal{A}$ denotes a finite 
alphabet (subsequently, the random variables will be identified with Ising 
spins, in which case $\mathcal{A}$ will denote the binary alphabet $\{-1,+1\}$). 
For $\sigma^M=\sigma_1\ldots \sigma_M$ denoting a particular symbol 
block of length $M>0$, the $M$-block Shannon entropy reads
\begin{align}
H[S^M]\equiv-\sum_{\sigma^M\in \mathcal{A}^M} {\rm Pr}(\sigma^M) \log_2({\rm Pr}(\sigma^M)) \label{eq:MblockShannonEntropy},
\end{align}
where ${\rm Pr}(\sigma^M)$ signifies the joint probability for blocks of $M$ 
consecutive symbols (i.e.\ spin orientations in the present context). 
In the above formula, the sum runs over all possible
blocks, i.e., combinations of of $M$ consecutive symbols from 
$\mathcal{A}$.

\subsection{Entropy density, excess entropy, and multi information for one-dimensional symbol sequences}
Using the $M$-block Shannon entropy, the asymptotic entropy density for a $1D$ system is given by 
\begin{equation}
h=\lim_{M\to\infty}H[S^M]/M\,.
\end{equation}

A sequence of finite-$M$ approximations $h(M)$ to the asymptotic entropy
density that typically converges faster than the expression above is provided by the 
\emph{conditional entropies}
\begin{align}
h(M)=H[S_M|S^{M-1}]= H[S^M] - H[S^{M-1}].\label{eq:conditionalEntropy}
\end{align}
As given above, $h(M)$ denotes the entropy of a single variable (spin) 
conditioned on a block of $M-1$ adjacent variables (spins). 
Thus, $h$ is the randomness that still remains, even after correlations 
over blocks of infinite length are accounted for.
Viewed as a function of block size, the finite-$M$ conditional entropies 
converge to the asymptotic value $h$ from above. Hence, at small length scales the system 
tends to look more random than it is in the limit $M\to\infty$.

For one-dimensional systems, there are three different but equivalent
expressions for the excess entropy. These are based on the convergence
properties of the entropy density, the subextensive part of the block entropy
in the limit of large block sizes, and, the mutual information between 
two semi-infinite blocks of variables, see Refs.\ \cite{Feldman2003,Feldman2008}.
Here, we focus on the definition of excess entropy $E$ that relates to the 
convergence properties of the entropy density in the form
\begin{align}
E=\sum_{M=1}^\infty [h(M)-h]. \label{eq:excessEntropy}
\end{align}
The conditional entropies $h(M)$ constitute upper bounds on the entropy rate,
allowing for an improving estimate of $h$ for increasing $M$. 
Thus, the individual terms in the sum comprise the entropy density 
overestimates 
on the level of blocks of finite length $M$. In total, the excess
entropy measures the area between $h(m)$ and the horizontal line at $h$.
As such, $E$ accounts for the randomness that is present at 
small lengths and that vanishes in the limit of large block sizes.

In $1D$ and in the limit of large block sizes, the \emph{multi information} is given 
by the first summand in Eq.\ \ref{eq:excessEntropy}, i.e.\
\begin{align}
I=h(1)-h. \label{eq:multiInformation}
\end{align}
Albeit $I$ is closely related to $E$ (this holds only in the limit of large
block sizes $M$; see Ref.\ \cite{Erb2004} for a more general discussion of the
multi information), it captures somewhat different 
characteristics regarding the convergence of the entropy density.
In this regard it measures the decrease of average uncertainty in the description
of the system by switching from the level of single variables (spins) statistics 
to the statistics attained as $M\to\infty$.

\begin{figure}[t!]
\begin{center}
\includegraphics[width=1.0\linewidth]{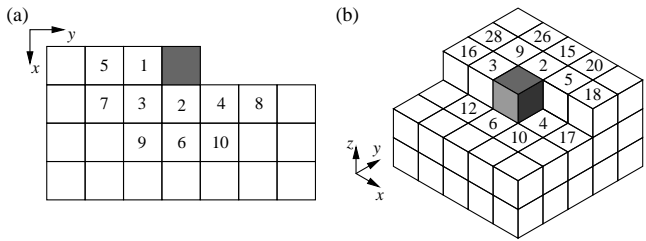}
\end{center}
\caption{\label{fig:nbTemplates}
Neighborhood templates for the definition of the contitional entropies (see Eq.\ \ref{eq:conditionalEntropy}) for 
(a) the $2D$ RBIM, and (b) the $3D$ RFIM. 
The target variable is represented by a shaded cell and the numbers within the 
cells reflect the order in which the sites are picked in order to compute the condititional entropies
(see text).
}
\end{figure}

\subsection{Extension to $d=2$ and $3$}

Following Ref.\ \cite{Feldman2003}, the most general approach designed to 
write the $2D$ entropy density as the entropy of a particular variable
conditioned on a block of neighboring variables (similar to the $1D$ case), 
uses so called neighborhood templates of a given size. 
In general, for a neighborhood template only spins in the same row left
of a ``target spin'' and the spins in all rows below the target spin are considered.
Fig.\ \ref{fig:nbTemplates}(a) shows such a finite-size template for a $2D$ system, 
covering an overall number of $25$ spins.
Similar to Ref.\ \cite{Feldman2003} we use the parameter $M$ in 
the conditional entropy (Eq.\ \ref{eq:conditionalEntropy}), to account for 
a successive addition of single sites from the neighborhood template shown in 
Fig.\ \ref{fig:nbTemplates}(a). 
Therein, the numbers within the cells
indicate the order in which the lattice sites are added to 
the $2D$ neighborhood template (from initial numerical experiments we 
found that it is not necessary to exceed $M=10$ ($M=6$) for the $2D$ RBIM ($3D$ RFIM)). 
In this regard, let the subscript $(\Delta x, \Delta y)$ denote the relative 
position of a variable $S$ in the neighborhood template, specifying its distance to 
the target variable in terms of site-to-site hops (in particular the 
target variable is labeled $S_{(0,0)}$). 
Then, the order in which the sites are picked from the template reflects the increasing 
Euclidean distance of the particular lattice site to the target site. 
Thereby, we follow the convention that if there is a draw regarding the distance
of two or more lattice sites, we add them to the neighborhood template from the 
top left to the bottom right, increasing the value of $\Delta y$ before $\Delta x$
(see the order of spins $7$ through $10$ in Fig.\ \ref{fig:nbTemplates}(a)).
As an example note that $h(6)=H[S_{(0,0)}|S_{(0,-1)} S_{(1,0)} S_{(1,-1)} S_{(1,1)} S_{(0,-2)}]$. 
We followed a similar approach in $3D$, where the corresponding neighborhood template is 
shown in Fig.\ \ref{fig:nbTemplates}(b), and where e.g.\ 
$h(4)=H[S_{(0,0,0)}| S_{(0,0,-1)} S_{(0,1,0)} S_{(-1,0,0)}]$.

Note that by the procedure discussed above, we disregarded 
the practice that if the interactions between the variables are of finite 
range, the neighborhood template only needs to be as ``thick'' as the interaction range \cite{Schlijper1990,Feldman2003}. 
For disordered model systems that exhibit competing interactions, 
nontrivial correlations between the spin degrees of freedom
might emerge that extend towards the intrinsic range of the interactions
(for a brief discussion of this issue and its implications, in particular
for the design of an efficient cluster algorithm for the fully frustrated 
Ising model, see Ref.\ \cite{Kandel1992}). Consequently, for the $2D$ RBIM
and $3D$ RFIM we considered it more adequate to include more than just the  
nearest neighbors for the construction of a neighbor template.

Subsequently, $2D$ and $3D$ configurations of spins are analyzed by parsing them 
into one dimensional sequences using the order of the 
spins as illustrated in Fig.\ \ref{fig:nbTemplates}. 
The resulting sequences can then be analyzed using Eqs.\
\ref{eq:conditionalEntropy} and \ref{eq:excessEntropy}, above.
From a practical point of view it will subsequently be necessary to truncate 
the sums in Eqs.\ \ref{eq:conditionalEntropy}, \ref{eq:excessEntropy} to 
a maximally feasible neighborhood size $M_{\rm max}$.


\section{Results}
\label{sect:results}
%

\begin{figure}[t!]
\begin{center}
\includegraphics[width=1.0\linewidth]{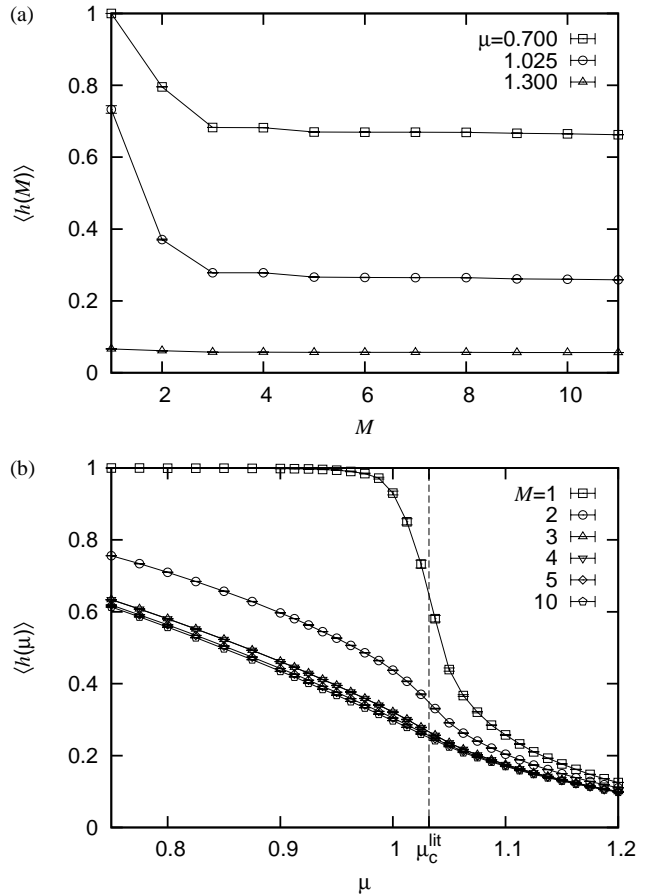}
\end{center}
\caption{\label{fig:2DRBIM_entropyDensity}
Results for the entropy density of the $2D$ RBIM with side length $L=384$.
(a) illustrates the convergence of the average entropy density $\langle h(M)\rangle$
as a function of the neighborhood size $M$ for three values of the disorder parameter $\mu$. 
(b) shows the average entropy density for different neighborhood sizes as a
function of the  disorder parameter $\mu$.
}
\end{figure}

\begin{figure*}[t!]
\begin{center}
\includegraphics[width=1.0\linewidth]{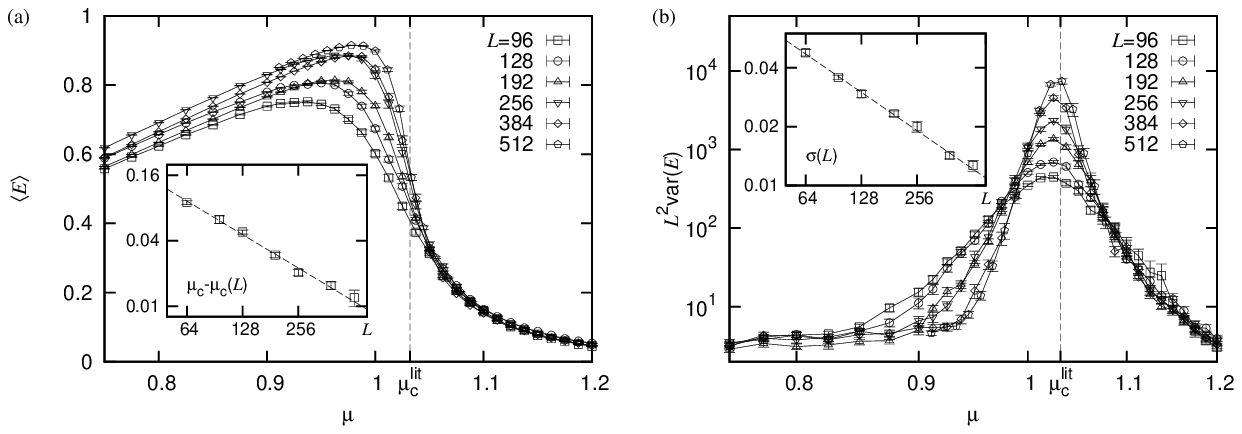}
\includegraphics[width=1.0\linewidth]{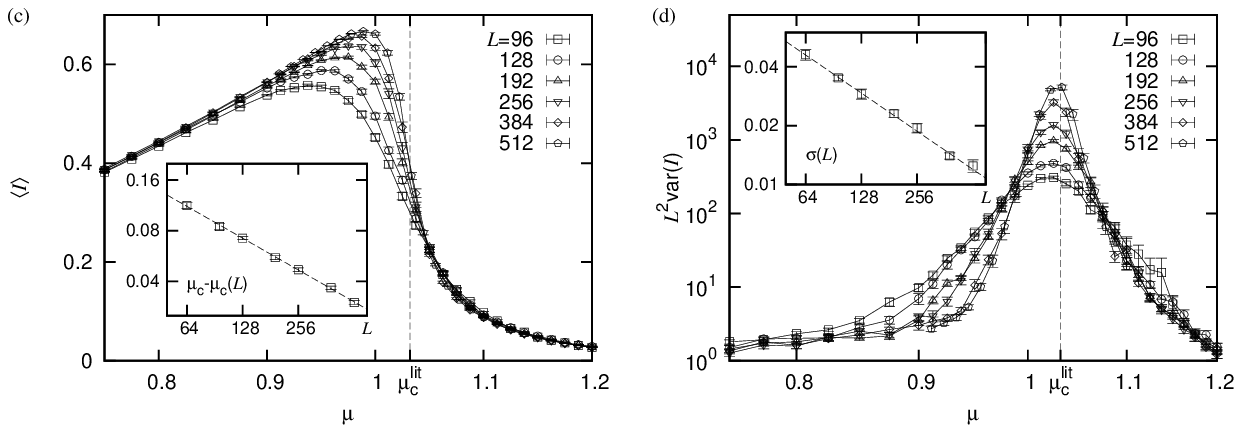}
\end{center}
\caption{\label{fig:2DRBIM}
Results for the $2D$ RBIM for different square lattices with side length up to $L=512$.
(a) The main plot shows the average excess entropy $\langle E \rangle$ as a 
function of  the mean $\mu$ of the bond disorder distribution (lines
are guides to the eyes only).
The inset illustrates the finite-size
scaling of the peak position, the line shows the result of a fit
to Eq.\ \ref{eq:scalingAssumption}. For all data-points
error bars are obtained via bootstrap resampling.
(b) The main plot shows the finite-size susceptibilities 
$\chi_E(L)=L^2 {\rm var}(E)$ 
associated to the excess entropy 
(lines are guides to the eyes only).
The inset signifies the finite-size scaling of the
peak width obtained from a fit of the data
points close to the peak to a Gaussian function,
the line shows the result of a fit
to Eq.\ \ref{eq:scalingFluct}.
(c),(d) are the same as (a),(b), respectively, but
for the multi-information $I$.
}
\end{figure*}
Here, our approach is somewhat different than that reported in Ref.\ \cite{Robinson2011}.
In that study, one particular realization of disorder for the Kaya-Berker model
was put under scrutiny at a particular value of the dilution parameter and for 
different temperatures (above as well as below the freezing temperature of the system).
For that particular disorder instance and for a given temperature, a single-spin flip 
Metropolis dynamics was used to generate independent spin configurations.
These spin configurations were then used to accumulate statistics for spin blocks
(specified by the utilized neighborhood template) that provide means to 
estimate conditional entropies on either local (i.e.\ lattice site dependent) or 
global scale, where the latter is a spatial average of the local conditional entropies.
As pointed out by the authors of Ref.\ \cite{Robinson2011}, it is important to accumulate 
spin-block statistics in a lattice site dependent manner so as to not overestimate the 
entropy of the system. A later average over the local observables yield the correct
thermodynamic entropy, as verified by the authors. 

However, note that here we work at $T=0$, aiming to characterize ground state phase transitions for 
disordered model systems in $2D$ and $3D$. I.e.\ for each realization of the 
disorder (and due to the particular disorder distributions used in the presented study),
there is one particular spin configuration that qualifies as a ground state (aside from a trivial 
degeneracy stemming from the global spin-flip symmetry of the 
energy function Eqs.\ \ref{eq:EA_hamiltonian}.
As a remedy, in order to collect statistics for blocks of spins we sweep the neighborhood
template over the full lattice (as one would do for pure systems, see Ref.\ \cite{Feldman2008}), 
thus averaging over different "local" configurations of the disorder. 
For one ground state and for a chosen block size $M$, this yields one spatially averaged estimate 
for the conditional entropy on the level of $M$-spin blocks.
The results are then averaged over many realizations of the disorder.
In doing so, we were interested whether one can observe a change in the structure
of the ground states by using the above information-theory inspired 
observables, and how this compares to phase transitions,
which were previously observed when using standard 
statistical-physics observables.

Furthermore,  
it is of interest to quantify the scaling behavior
of these observables in the vicinity of the phase transition. 
Note that
 we also performed additional simulations using the definition of 
$2D$ and $3D$ neighbor-templates as used in Ref.\ \cite{Schlijper1990} and found no qualitative difference 
to the results reported below.

\subsection{Results for the $2D$ RBIM}

\paragraph{Entropy density:}
As evident from Fig.\ \ref{fig:2DRBIM_entropyDensity}(a), the entropy density for
the $2D$ RBIM for $L=384$ exhibits a rapid convergence. 
In this regard, for a neighborhood size of $M=5$ the average entropy density appears 
to take its asymptotic value for the full range of considered parameters $\mu$.
However, for the numerical simulations at $L=384$ we considered the maximally feasible
neighborhood size $M_{\rm max}=10$. For smaller system sizes, $M_{\rm max}$
had to be adjusted to somewhat smaller values to assure that for a given system size $L$
the measurement of the conditional entropies at a given level $h(M)$ are based on sufficient statistics for 
the underlying $M$-block shannon entropies. In this regard, at $L=64$ we 
used $M_{\rm max}=6$.
The average entropy densities, restricted to neighborhood sizes $M=1$ through $10$,
are shown in Fig.\ \ref{fig:2DRBIM_entropyDensity}(b). It can be seen that for 
small (large) values of $\mu$ the entropy density converges to a comparatively large (small)
value representing the disordered (i.e.\ SG ordered) and ordered (i.e.\ ferromagnetic) phase, respectively. 
Especially the curve corresponding to $M=1$ exhibits a steep decrease in the
interval $\mu\in[1,1.1]$. Moreover, the fluctuations ${\rm var}(h(\mu))\equiv\langle h(\mu)^2\rangle-\langle h(\mu)\rangle^2$ 
of the entropy density are peaked at $\mu\approx 1.025$ (not shown). As it appears, this 
parameter value is close to the asymptotic critical point $\mu_{\rm c}^{\rm lit}=1.030(2)$ that indicates the
$T=0$ SG to FM transition for the RBIM \cite{McMillan1984,Melchert2009} in the limit of
large system sizes.

\paragraph{Excess entropy:}
A finite-size scaling analysis of the system size dependent
peak position of the average excess entropy $\langle E\rangle$ (see Fig.\ \ref{fig:2DRBIM}(a)) 
was performed in the following way:
polynomials of order $5$ were fitted to the data curves 
at different system sizes $L$ in order to obtain an estimate
$\mu_{{\rm c},i}(L)$ of the peak position. Thereby, the index $i$ 
labels independent estimates of the peak positions as 
obtained by bootstrap resampling \cite{practicalGuide2009}. 
For the analysis, we considered
$40$ bootstrap data sets, e.g.\ resulting in the final estimate
$\mu_{\rm c}(L=256)=0.975(1)$. Anticipating the scaling form 
\begin{eqnarray}
\mu_{\rm c}(L)=\mu_{\rm c} - a L ^{-b}, \label{eq:scalingAssumption}
\end{eqnarray}
see inset of Fig.\ \ref{fig:2DRBIM}(a),
and considering the fit interval $L\in[64,512]$ yields the fit parameters 
$\mu_{\rm c}=0.995(5)$, $b=1.0(1)$ and $a=O(1)$ for a reduced chi-square $\chi^2/{\rm dof}=1.19$ (${\rm dof}=4$).
Note that as $L\to\infty$, the location of the peak disagrees with 
the location $\mu_{\rm c}^{\rm lit}=1.031(2)$ of the $T=0$ SG to FM transition.
The latter estimate was obtained from an analysis of the binder-parameter
for the GS magnetization \cite{Melchert2009}.
Earlier simulations, using a transfer-matrix approach at finite temperature and extrapolated 
to $T=0$, report on $\mu_{\rm c}^{\prime {\rm lit}}=1/r_{\rm c}=1.04(1)$ \cite{McMillan1984}.

A similar scaling analysis for the location of the peaks related to the finite-size
susceptibilities $\chi_E(L)\equiv L^2 {\rm var}(E)$ by means of Gaussian
fit-functions results in the 
estimate $\mu_{\rm c}=1.029(1)$. Further, the width of the Gaussian 
fit-function obeys the scaling 
\begin{eqnarray}
\sigma(L)=aL^{-b}, \label{eq:scalingFluct}
\end{eqnarray}
where $a=O(1)$ and $b=0.65(2)$ ($\chi^2/{\rm dof}=0.45$, ${\rm dof}=5$).
Note that the inverse of the latter fit parameter reads $1/b=1.54(5)$
and is thus strikingly close to the critical exponent $\nu^{\rm lit}=1.49(7)$ \cite{Melchert2009} 
($\nu^{\prime {\rm lit}}=1.42(8)$ \cite{McMillan1984}) 
that characterizes the $T=0$ SG to FM transition.

\begin{figure}[t!]
\begin{center}
\includegraphics[width=1.0\linewidth]{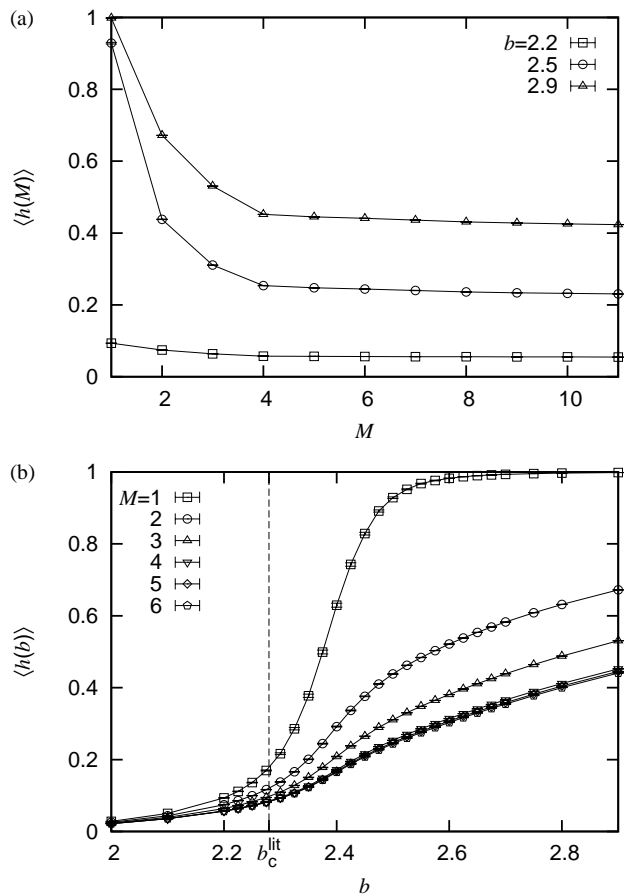}
\end{center}
\caption{\label{fig:3DRFIM_entropyDensity}
Results for the entropy density of the $3D$ RFIM with side length $L=64$.
(a) illustrates the convergence of the average entropy density 
$\langle h(M)\rangle$ as a function of the neighborhood size $M$ 
for three values of the field strength $b$. 
(b) shows the average entropy density for different neighborhood sizes as 
a function of the  field strength $b$.}
\end{figure}

\begin{figure*}[t!]
\begin{center}
\includegraphics[width=1.0\linewidth]{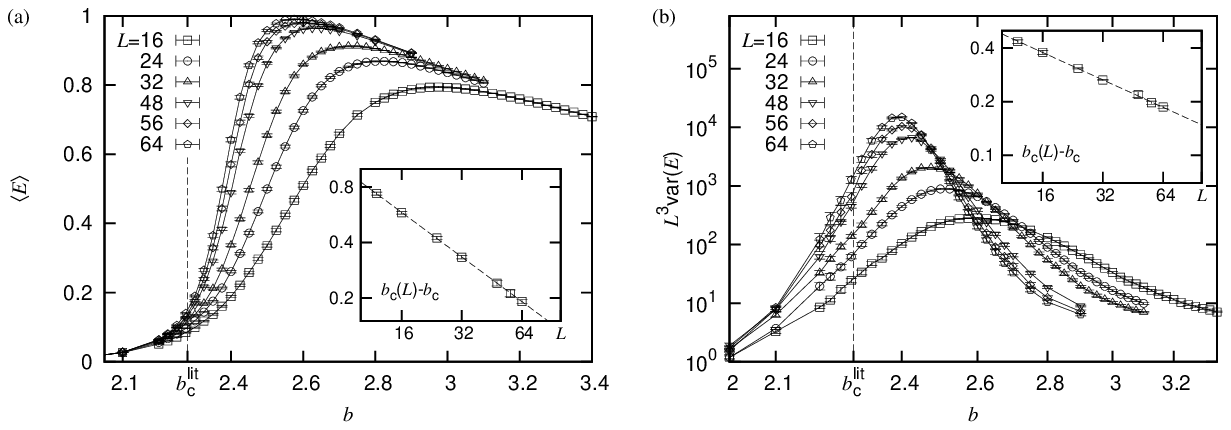}
\includegraphics[width=1.0\linewidth]{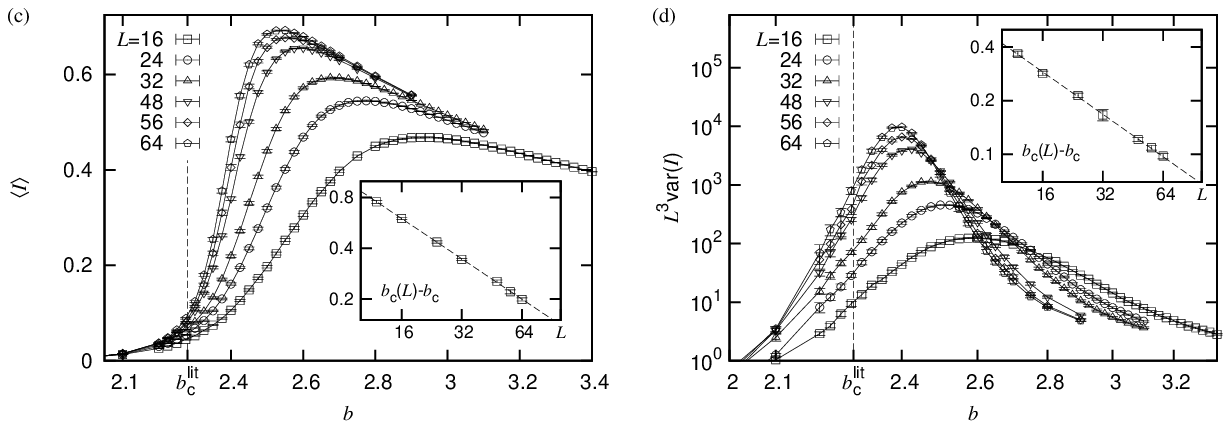}
\end{center}
\caption{\label{fig:3DRFIM}
Finite-size scaling analysis for the $3D$ RFIM on cubic lattices with side length up to $L=64$.
(a) the main plot shows the average excess entropy $\langle E \rangle$ 
as a function of 
the field strength $b$. (For this and the other main plots,
lines are guides to the eyes only).
The inset illustrates the finite-size
scaling of the peak position, the line, as for the other
insets, shows the result of a fit
to Eq.\ \ref{eq:scalingAssumptionB}. For all data-points
 error bars are obtained via bootstrap resampling,
(b) the main plot shows the finite-size susceptibilities $\chi_E(L)=L^3 {\rm var}(E)$ 
associated to the excess entropy and the inset indicates the finite-size scaling of the
associated peak location.
(c),(d) are the same as (a),(b), respectively, but for the multi-information $I$.
}
\end{figure*}
\paragraph{Multi-information:}
A scaling analysis of the system size dependent
peak position displayed by the average multi-information $\langle I\rangle$ (see Fig.\ \ref{fig:2DRBIM}(c)) 
according to the scaling assumption Eq.\ \ref{eq:scalingAssumption} in the range $L\in[64,512]$ yields
the fit-parameters $\mu_{\rm c}=1.026(3)$, $a=O(1)$, and $b=0.63(4)$, i.e.\ $1/b=1.6(1)$ ($\chi^2/{\rm dof}=0.20$, ${\rm dof}=4$).
Neglecting the smallest system size, i.e.\ restricting the fit interval to $L\in[96,512]$, leads to 
the estimate $\mu_{\rm c}=1.031(6)$ which is even closer to the known critical point. 

Further, an analysis of the position and width of the peaks related
to the finite-size fluctuations $\chi_I\equiv L^2 {\rm var}(I)$ by means 
of Gaussian fit-functions results in the additional 
estimate $\mu_{\rm c}=1.032(5)$ (see Fig.\ \ref{fig:2DRBIM}(d)). As above, the widths $\sigma(L)$ obay the 
scaling form of Eq. \ref{eq:scalingFluct}, where a fit yields $a=O(1)$ and $b=0.65(2)$, i.e.\ $1/b=1.54(5)$ 
($\chi^2/{\rm dof}=0.32$, ${\rm dof}=5$; inset of Fig.\ \ref{fig:2DRBIM}(d)).

Note that here, both estimates of $\mu_{\rm c}$ and both estimates of the exponent $1/b$ 
are reasonably close to those found earlier for the $2D$ RBIM. Hence, this
might indicate that in comparison to the full excess entropy, the multi-information is more sensitive to structural 
changes at the $T=0$ order-to-disorder transition in the $2D$ RBIM.

\paragraph{Geometric properties of the GSs:}
Upon increasing the value of the disorder parameter from $\mu=0$ to $\mu>\mu_c$, it
is possible to identify ferromagnetic clusters of spins with increasing size.
A finite size scaling analysis of the relative size of the largest and second largest ferromagnetic
clusters of spins for the independent GSs can be utilized to characterize the
$T=0$ SG to FM transition in the $2D$ RBIM. E.g.\ as reported in Ref.\ \cite{Melchert2009},
the relative size of the largest ferromagnetic cluster scales as 
$\langle M_1 \rangle\propto L ^{-\beta/\nu} f[(\mu-\mu_{\rm c})L^{1/\nu}]$, where
a data collapse (for system sizes $L=24\ldots 64$) yields the scaling parameters $\mu_{\rm c}^{\rm fc}=1.032(2)$, $\nu^{\rm fc}=1.49(4)$, and $\beta^{\rm fc}=0.039(4)$.
Note that the results obtained for the multi-information
is also in excellent agreement with those obtained from an analysis of the largest cluster size,
which constitutes an observable that links to the geometric properties of the GS spin configurations.

\subsection{Results for the $3D$ RFIM}

\paragraph{Entropy density:}
As can be seen from Fig.\ \ref{fig:3DRFIM_entropyDensity}(a), the entropy density for
the $3D$ RFIM (the figure shows the data for $L=64$) convergence rapidly. 
I.e.\ for a neighborhood size of $M=6$ the average entropy density appears 
to take its asymptotic value for the full range of considered field strengths $b$.
During the numerical simulations, carried out on cubic systems of side length $L=8$ 
through $64$, we thus considered the maximally feasible neighborhood size $M_{\rm max}=6$. 
The average entropy densities for $M=1\ldots 6$ are shown in Fig.\ \ref{fig:3DRFIM_entropyDensity}(b). 
For the $3D$ RFIM the ordered, i.e., ferromagnetic (FM)) and disordered, i.e.,
 paramagnetic (PM) phases 
are located at small and large values of the disorder parameter
$b$, respectively.
As discussed above, the curve corresponding to $n=1$ exhibits a steep decrease in the
interval $b\in[2.2,2.5]$, which already allows to shed some light on where the characteristics
of the model at $T=0$ change from ferromagnetic to paramagnetic. 
In a previous study, using the finite-size scaling of
standard physical quantities,
the critical strength of the random field at which the $T=0$ FM to PM transition
takes place was found to be $b_{\rm c}^{\rm lit}=2.28(1)$ \cite{hartmann2001}.

\paragraph{Excess entropy:}
The scaling analysis of the system size dependent
peak position of the average excess entropy $\langle E\rangle$ (see Fig.\ \ref{fig:3DRFIM}(a)) 
was performed similar to the $2D$ RBIM above.
Assuming that the system size dependent peak locations $b_{\rm c}(L)$ exhibit a scaling of the form 
\begin{eqnarray}
b_{\rm c}(L)=b_{\rm c} - a L ^{-c}, \label{eq:scalingAssumptionB}
\end{eqnarray}
see inset of Fig.\ \ref{fig:3DRFIM}(a),
yields the fit parameters 
$b_{\rm c}=2.38(1)$, $c=0.81(2)$ and $a=O(1)$ ($\chi^2/{\rm dof}=0.05$, ${\rm dof}=4$).
Note that as $L\to\infty$, the location of the peak extrapolates 
to a value that is close by the critical point $b_{\rm p}^{\rm lit}=2.32(1)$, describing the transition of the
probability for a simultaneous spanning of up and down spin domains from $1$ to $0$, see Ref.\ \cite{Seppala2002}.
A similar scaling analysis for the location of the peaks related to the finite-size
susceptibilities $\chi_E(L)\equiv L^2 {\rm var}(E)$, again by means of 
polynomials of order $5$, results in the estimate $b_{\rm c}=2.23(1)$, see inset of \ref{fig:3DRFIM}(b). 
As it appears, the estimates of $b_{\rm c}$ obtained from the excess entropy and its fluctuation do not 
match up well.

\paragraph{Multi-information:}
A finite-size scaling analysis of the system size dependent
peak position shown by the average multi-information $\langle I\rangle$ (see Fig.\ \ref{fig:3DRFIM}(c)) 
according to the scaling assumption Eq.\ \ref{eq:scalingAssumptionB} in the range $L\in[12,64]$ yields
the fit-parameters $b_{\rm c}=2.34(1)$, $a=O(1)$, and $c=0.79(3)$ ($\chi^2/{\rm dof}=0.22$, ${\rm dof}=4$).
Further, an analysis of the peak location for the related
finite-size fluctuations $\chi_I\equiv L^3 {\rm var}(I)$ 
results in the estimates $b_{\rm c}=2.30(1)$, $a=O(1)$ and $c=0.80(7)$ 
($\chi^2/{\rm dof}=0.26$, ${\rm dof}=5$); see Fig.\ \ref{fig:3DRFIM}(d).

Note that here, both estimates of $b_{\rm c}$ and both estimates of the exponent $c$ 
match up well.
Also, the numerical estimates $b_{\rm c}$ agree with the numerical value of $b_{\rm p}^{\rm lit}$ within
errorbars and the numerical value of $1/c=1/0.80(7)=1.3(1)$ (obtained by means of the above
analyses) is in good agreement with the correlation length exponent $\nu^{\rm lit}=1.3(1)$ reported in Ref.\ \cite{hartmann2001}.
Further, they are in good agreement with the numerical values
to which the scaling analysis for the percolation criterion in Ref.\ \cite{Seppala2002} extrapolates to.
Again, these results indicate that the multi-information is very sensitive to structural 
changes in GS spin configurations at the $T=0$ order-to-disorder transition.

\subsection{The $T=0$ phase transition in purely information theoretic coordinates}

\begin{figure}[t!]
\begin{center}
\includegraphics[width=1.0\linewidth]{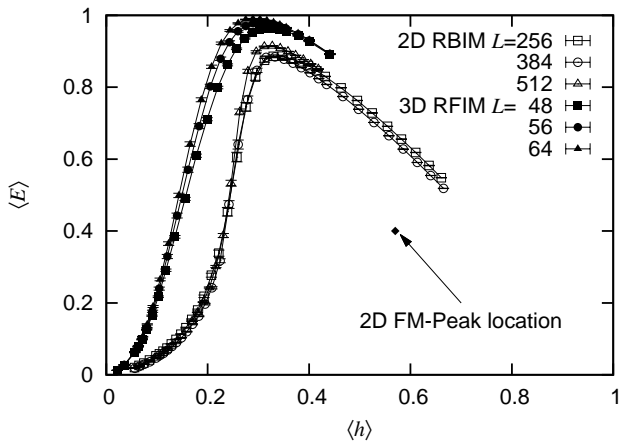}
\end{center}
\caption{\label{fig:entropyComplexity}
Complexity-entropy diagram for the $2D$ RBIM and $3D$ RFIM for 
different system sizes $L$. 
Note that for the complexity-entropy curve for the $2D$ Ising FM,
the corresponding peak is located at entropy
density $\langle h \rangle \approx 0.57$ with a 
peak height of $\langle E\rangle \approx 0.4$, see Ref.\ \cite{Feldman2008}.}
\end{figure}

In the above analyses the information theoretic coordinates entropy density $h$ and 
excess entropy $E$, giving measures of randomness and complexity, respectively, 
were studied as a function of a model specific parameter.
So as to facilitate a comparison of different models 
\emph{complexity-entropy diagrams} 
are of great use \cite{Crutchfield1989}. A survey of complexity-entropy 
diagrams for different model systems
can be found in Ref.\ \cite{Feldman2008}. The complexity-entropy 
relationship for the $2D$ RBIM and 
$3D$ RFIM for different system sizes are shown in 
Fig.\ \ref{fig:entropyComplexity}.
As evident from the figure, the curves for both models exhibit 
similar features.
I.e.\ they have an isolated peak at a particular value of $h$. This is 
similar to the  complexity-entropy curve for the $2D$ Ising FM,
where the corresponding peak is located at entropy
density $\langle h \rangle \approx 0.57$ with a 
peak height of $\langle E\rangle \approx 0.4$, see Ref.\ \cite{Feldman2008}. 
In order to illustrate finite-size effects for the complexity-entropy diagram, 
three different system sizes for both models are shown. A scaling analysis of
the finite-size peak locations indicates that for the $2D$ RBIM ($3D$ RFIM) the peak 
shifts to the entropy density value $h_{\rm c}=0.26(1)$ 
($h_{\rm c}=0.107(1)$) as $L\to\infty$.
Comparing with the result of the FM mentioned before, one
can say that for the $2D$ RBIM and the $3D$ RFIM, the order-disorder
transition appears at smaller entropy but is connected with
a higher complexity. Note, for the considered models, 
the complexity-entropy diagram contains the FM phase 
in the parameter range $h\in[0,h_{\rm c}]$. The SG phase for the $2D$ RBIM 
(PM phase for the $3D$ RFIM) is found for $h\in[h_{\rm c},1]$. 

\section{Discussion}
\label{sect:discussion}

Regarding the excess entropy, conceptually similar analyses carried out on spin 
configurations for the $2D$ Ising FM, 
reported in Ref.\ \cite{Feldman2008}. The respective study concluded that the excess entropy is
peaked at a temperature $T\approx 2.42$ in the paramagnetic (i.e.\ disordered) phase 
above the true critical temperature $T_c=2.269$.
Qualitatively similar results on the mutual information for the $2D$ Ising FM 
(and more general classical $2D$ spin models)
were recently presented in Ref.\ \cite{Wilms2011}. There, the authors conclude
that the mutual information (equivalent to the excess entropy for $1D$ systems; 
see Ref.\ \cite{Crutchfield2003}) reaches a peak-value in the paramagnetic
phase close to the system parameter $K=J/k_BT\approx0.41$ (for $J=k_B=1$ 
this corresponds to $T\approx 2.44$), again in disagreement with the critical 
point of the underlying model. 
Further, the authors note that the true critical point appears to coincide
with the inflection point where the first derivative of the excess entropy tends
to minus infinity in the thermodynamic limit (however, the authors present no 
systematic analysis of this observation).
Hence, it is not too surprising that a finite-size scaling analysis for the peaks
of the bare excess entropy does not directly allow to identify the critical point of 
the model considered here. Similar to the previous studies for the Ising FM, 
we here find that the excess entropy assumes an isolated peak at a parameter-value 
located slightly below the true critical point in the disordered phase. 
However, a scaling analysis reveals that in the thermodynamic limit, the peak of the 
related finite-size fluctuations is located right at the critical point.

The multi-information was introduced by Erb and Ay in Ref.\ \cite{Erb2004}, 
where the authors considered the multi-information 
to characterize spin-configuration for the $2D$ Ising FM in the thermodynamic limit by analytic means. 
Among other things, the authors conclude that the multi-information exhibits an isolated global maximum right 
at the critical temperature (see Theorem 3.3 of Ref.\ \cite{Erb2004}). 
Here, we find that in contrast to the excess entropy and in qualitative agreement with analytic results
for the $2D$ Ising FM,
the peak of the multi-information tends towards the critical points reported in the literature as $L\to \infty$. 
However, the critical point to which the effective, system size dependent critical points related 
to the peaks of the excess entropy for the $3D$ RFIM converges is in good agreement with 
the precise location that corresponds to a particular percolation transition for the respective
spin model. Nevertheless, given the so far achieved numerical
accuracy, it is still not clear to us whether these two transition
points are really distinct. Anyway, both observables studied in this work 
basically turned out to be useful information-theoretic measures that 
might be used to distinguish
the ordered and disordered phase of frustrated model systems, 
as e.g.\ the $2D$ RBIM and $3D$ RFIM.
Further, the finite size scaling of these observables allows to estimate 
critical points and exponents that are in good agreement with the 
critical properties reported in the literature.

Given these promising results for systems exhibiting quenched
disorder, it would be of particular interest to apply these
methods to structural glasses \cite{binder2011}, 
where a simple way of analyzing snapshots
of configurations is still missing \cite{biroli2008}.

\begin{acknowledgments}
We thank F.\ Krzakala for valuable discussions and for pointing out Ref.\ \cite{biroli2008}.
OM acknowledges financial support from the DFG (\emph{Deutsche Forschungsgemeinschaft})
under grant HA3169/3-1.
The simulations were performed at the HPC Cluster HERO, located at 
the University of Oldenburg (Germany) and funded by the DFG through
its Major Instrumentation Programme (INST 184/108-1 FUGG) and the
Ministry of Science and Culture (MWK) of the Lower Saxony State.
\end{acknowledgments}


\bibliographystyle{unsrt}
\bibliography{lit_entropyComplexity.bib}

\end{document}